\documentclass[11pt]{article}
\usepackage{epsfig}
\usepackage[usenames]{color}
\usepackage{graphicx}
\usepackage{amsmath}
\newcommand{\sbar}[1]{\ooalign{\hfil/\hfil\crcr$#1$}}

\def\la{\langle}\def\ra{\rangle}
\def\be{\begin{eqnarray}}\def\ee{\end{eqnarray}}
\def\lsim{\mathrel{\rlap{\lower3pt\hbox{\hskip1pt$\sim$}}
     \raise1pt\hbox{$<$}}} 
\def\gsim{\mathrel{\rlap{\lower3pt\hbox{\hskip1pt$\sim$}}
     \raise1pt\hbox{$>$}}} 
\def\le{ \begin{array}{ll}}\def\re{\end{array}}
\def\mn{m_{N^{+}}}\def\mm{m_{N^{-}}}\def\mpi{m_{\pi}}
\def\lear{ \left( \begin{array}{cc}}\def\rear{\end{array} \right)}

\def\mpi{m_{\pi}}
\def\mn{m_{N_{+}}}\def\mm{m_{N_{-}}}
\def\le{ \left( \begin{array}{cc}}\def\re{\end{array} \right)}

\def\mpi{m_{\pi}}\def\MeV{\textnormal{MeV}}
\def\bi{\bibitem}

\renewcommand{\theequation}{\arabic{section}.\arabic{equation}}

\renewcommand{\thefootnote}{\fnsymbol{footnote}}
\begin{document}
\hfill \vbox{\hbox{}}
\begin{center}{\Large\bf Dilaton-Limit Fixed Point\\ in Hidden Local Symmetric Parity Doublet Model}\\[1.5cm]
{ Won-Gi Paeng\footnote{\sf e-mail: wgpaeng0@hanyang.ac.kr}
}\\
{\em Department of Physics, Hanyang University, Seoul 133-791, Korea}

{ Hyun Kyu Lee\footnote{\sf e-mail: hyunkyu@hanyang.ac.kr}
}\\
{\em Department of Physics, Hanyang University, Seoul 133-791, Korea \&
\\Asia Pacific Center for Theoretical physics, Pohang, Gyeongbuk 790-784, Korea}

{ Mannque Rho\footnote{\sf e-mail: mannque.rho@cea.fr }
}\\
{\em Institut de Physique Th\'eorique, CEA Saclay, 91191 Gif-sur-Yvette c\'edex, France \&
\\Department of Physics, Hanyang University, Seoul 133-791, Korea}

{ Chihiro Sasaki\footnote{\sf e-mail: sasaki@fias.uni-frankfurt.de}
}\\
{\em Frankfurt Institute for Advanced Studies,
D-60438 Frankfurt am Main, Germany}

\end{center}
\vspace{0.5cm}
\centerline{\today}
\vspace{0.5cm}
\begin{center}{\Large\bf Abstract}
\end{center}


%
%
%
%

We study nucleon structure with positive and negative parities using a parity doublet model endowed with hidden local symmetry (HLS) with the objective to probe dense baryonic matter. The model -- that we shall refer to as ``PDHLS model" for short -- allows a chiral-invariant mass of the nucleons unconnected to spontaneously broken chiral symmetry which comes out to be $m_0 \sim 200$ MeV at tree level from fitting to the decay width of the parity doubler, N(1535), to nucleon-pion and nucleon axial coupling $g_A=1.267$. The
presence of a substantial $m_0$ that remains non-vanishing at
chiral restoration presents a deep issue for the origin of
the nucleon mass as well as will affect nontrivially the
equation of state for dense baryonic matter relevant for
compact stars. We construct a chiral perturbation theory
at one-loop order and explore the phase structure of the model
using renormalization group equations. We find a fixed point
that we identify with the  ``dilaton limit" at which the HLS
vector mesons decouple from the nucleons. We suggest that cold baryonic system will flow to this limit either before or at reaching the vector manifestation fixed point of hidden local symmetry theory as density increases toward that of chiral
restoration.



\vfill

\pagebreak
\setcounter{footnote}{0}
\renewcommand{\thefootnote}{\arabic{footnote}}
\section{Introduction}
In the effort to decipher what happens when hadronic matter is
compressed to high density as gravity does in compact stars
(or to high temperature as in relativistic heavy ion
collisions)\footnote{In this paper, we will be mainly concerned
with density although we will make references to temperature.},
hidden local symmetry (HLS)~\cite{HLStree,hls} promises to be a
powerful and predictive theoretical tool, hitherto more or less
unexploited. Hidden local symmetry naturally arises when
nonlinear sigma  (NL$\sigma$) model is extended to the energy
scale commensurate with the mass of the vector ($\rho$,
$\omega$) mesons. It can be taken as emergent from the current
algebra scale~\cite{Son} or reduced from string theory via
holography~\cite{SS,HMY}.

HLS makes certain remarkable predictions that are both simple
and unanticipated by other approaches available in the
literature. While their validity rests on certain assumptions
that await confirmations or refutations by QCD proper or
experiments, if valid, their consequences on hadronic matter
under extreme conditions could be enormous.

There are two particularly notable predictions that we are
concerned with here,  both of which have not been made in other
models. One is that as the hadronic matter approaches the
extreme condition -- either high temperature or high density --
at which a phase transition takes place from broken to restored
chiral symmetry, the vector meson mass $m_V$ (where $V=\rho,
\omega$) should scale as~\cite{hls} \be m_V^*/m_V\approx
\la\bar{q}q\ra^*/\la\bar{q}q\ra \ee where the asterisk denotes
in medium -- temperature $T$ or density $n$ -- and $q$ stands
for the chiral quark (massless in the chiral limit). More
conjecturally, this relation has been extended to other
light-quark mesons, leading to BR scaling~\cite{BR91}. One can
also write down a similar scaling for the nucleon~\cite{hkr} if
one takes the nucleon in the standard (or ``naive") assignment,
that is, anchored on the assumption that the nucleon mass is
entirely generated dynamically, i.e., by spontaneous breaking
of chiral symmetry.

Another striking prediction that is also quite distinctive from
others is on the short-distance properties in nuclear
interactions in dense medium. If one implements the trace
anomaly of QCD in terms of dilatons into the HLS
Lagrangian~\cite{LR}\footnote{Note that introducing scalar
excitations into the NL$\sigma$ model and equivalently into HLS
has been problematic. Here we rely on the locking of scale
symmetry with chiral symmetry as discussed in \cite{LR}.} that
we will refer to as dHLS, and if one makes the reasonable
assumption that what is called ``dilaton limit"~\cite{beane}
is simulating the approach to chiral restoration in
density as will be argued below, then the prediction is that
the vector-meson coupling to the nucleon $g_{VNN}$ should get
suppressed at high density~\cite{Sasaki}. As a consequence, we
have that as density increases, (1) the repulsive core produced by $\omega$-meson exchanges gets suppressed; (2) the tensor force contributed by the $\rho$ exchange gets also suppressed. These two effects are expected to have a large impact on the EoS for compact stars.

We should, however, note that some, if not all, of the
predictions mentioned above that involve fermionic, i.e.,
nucleonic degrees of freedom, can be modified by the
possibility that part of the nucleon mass may not be generated
dynamically, as for instance in the case of mirror assignment
for the nucleons~\cite{DeTar,mirror}. This would mean that a
part of the mass, say, $m_0$, would not vanish up to the chiral
restoration point. Such a model would substantially modify or even invalidate the BR scaling for the nucleon although it may leave intact the one for
mesons.

In fact such a structure is found to arise when skyrmions are
put on crystal lattice to simulate dense matter. This is a
picture of dense baryonic matter for large $N_c$ at which the
crystal structure is justified with the nucleon mass going
proportional to $N_c$. What one finds~\cite{LPR} is that a
skyrmion matter at a lower density with $\la\bar{q}q\ra^*\neq
0$ and $f_\pi^*\neq 0$ undergoes a phase transition at a higher
density $n=n_{1/2}> n_0$ to a matter composed of half-skyrmions
with $\la\bar{q}q\ra^*=0$ and $f_\pi^*\neq 0$~\footnote{
  The generalized pion decay constant $f_\pi^\ast$ can be saturated
  by not only $\bar{q}q$ but also some higher-dimension operators,
  such as a four-quark, which could remain condensed giving
  $f_\pi^\ast \neq 0$ whereas $\langle\bar{q}q\rangle$ is suppressed
  in a phase at high density.
  We suggest that the half-skyrmion phase is such an example.
}. In this
half-skyrmion phase, the in-medium nucleon mass scales
proportionally to $f_\pi^*$ which does not drop appreciably up
to the critical density. This suggests that effectively, there
is a non-vanishing $m_0$. This means that the in-medium nucleon
mass does not follow the scaling of light-quark mesons. A
notable consequence of this structure is that above $n_0$ at,
say, $(1.3 -2.0)n_0$, there is a significant change in the
structure of nuclear tensor forces -- and hence in the equation of state of compact-star matter. In the presence of both
pion and $\rho$,  the tensor forces are given by contributions
with opposite signs from the pion exchange and $\rho$ exchange.
In medium, however, the $\rho$ contribution -- which is
repulsive -- tends to suppress the attraction due to the pion
exchange, stiffening the spin-isospin interactions~\cite{BR
tensor} -- which is consistent with nature up to nuclear matter
density. However if the scaling of the nucleon mass is modified
substantially from Brown-Rho scaling by the presence of a large
$m_0$, then it turns out that as mentioned above, the $\rho$ tensor gets suppressed at some high density, and the pion tensor starts dominating.
This feature is expected to have a drastic impact on the EoS,
particularly on the symmetry energy $E_{sym}$, relevant to
compact stars~\cite{LPR}.

The objective of this paper is to expose the role of the chiral
invariant mass $m_0$ in the nucleon in the framework of an
effective field theory anchored on hidden local symmetry.  For
this purpose, we construct a parity-doublet model for baryons
in a hidden local symmetric setting  that we shall refer to as ``parity-doublet HLS (PDHLS) model" and study what value of
$m_0$ is allowed by nature. We should stress that at this
point, we do not know whether this model captures fully the
physics of the half-skyrmion phase where a non-zero $m_0$ is
indicated. This is an issue to be addressed further. In this
paper, we will restrict ourselves to phenomenology in the
vacuum, namely, at $T=n=0$, relegating in-medium properties to
a later publication. Our estimate of $m_0$ will be at tree
order. We will however look at one-loop RGEs for two-point and
three-point functions and establish that the dilaton limit
taken in \cite{Sasaki} to approach mended symmetry corresponds to a fixed point in the one-loop RGE flow in the standard (or ``naive") assignment and also in the mirror assignment for the baryons . The plan of our paper is as follows: we introduce our Lagrangian with parity doublers in
Section~\ref{sec:hls} and deduce an $m_0$ from the known
phenomenology in matter-free space in Section~\ref{sec:m0}.
Analysis of the RGEs and the phase structure with the PDHLS model is made in
Section~\ref{sec:rge}. A summary and conclusions are given in
Section~\ref{sec:conclusions}. Detailed expressions are
summarized in Appendices.

\setcounter{equation}{0}
\section{Hidden Local Symmetry in Parity Doublet Model }
\label{sec:hls}

In this section we give a brief introduction of a nonlinear chiral
Lagrangian based on hidden local symmetry (HLS)~\cite{HLStree} and
introduce parity doubled nucleons~\cite{DeTar,mirror}. Here and in what follows, we consider a system with $N_f=2$.

The 2-flavored HLS Lagrangian is based on
a $G_{\rm{global}} \times H_{\rm{local}}$ symmetry,
where $G_{\rm global}=[SU(2)_L \times SU(2)_R]_{\rm global}$
is the chiral symmetry and
$H_{\rm local}=[SU(2)_V]_{\rm local}$
is the HLS.
The whole symmetry $G_{\rm global}\times H_{\rm local}$
is spontaneously broken to a diagonal $SU(2)_V$.
The basic quantities are
the HLS gauge boson, $V_\mu$,
and
two matrix valued variables $\xi_L$, $\xi_R$,
which are combined in a
$2 \times 2$ special-unitary matrix
$U = \xi_L^\dagger \xi_R$.
The transformation property of $U$ under the chiral symmetry is
given by
\begin{equation}
U \to g_L U g_R^\dagger\,,
\end{equation}
with $g_{L,R} \in \left[ SU(2)_{L,R}\right]_{\rm global}$.
The variables $\xi$s transform as
\begin{equation}
\xi_{L,R} \to h\,\xi_{L,R}\,g_{L,R}^\dagger\,,
\end{equation}
with $h \in \left[ SU(2)_V\right]_{\rm local}$, and are parameterized as
\begin{equation}
\xi_{L,R}=e^{i\sigma/{F_\sigma}}e^{\mp i\pi/{F_\pi}}\,,
\end{equation}
where $\pi = \pi^a T_a$ denotes the pseudoscalar
Nambu-Goldstone (NG)
bosons associated with the spontaneous symmetry breaking of
$G_{\rm{global}}$ chiral symmetry,
and $\sigma = \sigma^a T_a$ denotes
the NG bosons associated with
the spontaneous breaking of $H_{\rm{local}}$.
The $\sigma$ is absorbed into the HLS gauge boson through
the Higgs mechanism and the gauge boson acquires its mass.
$F_\pi$ and $F_\sigma$ are the decay constants
of the associated particles.
The HLS gauge field transforms as
\begin{equation}
V_\mu \to ih\partial_\mu h^\dagger + hV_\mu h^\dagger\,.
\end{equation}

The fundamental objects are the Maurer-Cartan 1-forms
defined by
\begin{eqnarray}
\hat{\alpha}_{\perp }^{\mu}
&=& \frac{1}{2i}\left[ D^\mu\xi_R \cdot \xi_R^{\dagger}
{}- D^\mu\xi_L \cdot \xi_L^{\dagger} \right]\,,
\nonumber\\
\hat{\alpha}_{\parallel}^{\mu}
&=& \frac{1}{2i}\left[ D^\mu\xi_R \cdot \xi_R^{\dagger}
{}+ D^\mu\xi_L \cdot \xi_L^{\dagger} \right]\,,
\end{eqnarray}
which transform homogeneously:
\begin{equation}
\hat{\alpha}_{\perp,\parallel}^\mu
\to h\, \hat{\alpha}_{\perp,\parallel}^\mu\, h^\dagger\,.
\end{equation}
The covariant derivatives of $\xi_{L,R}$ are given by
\begin{eqnarray}
&&
D_\mu \xi_L
 = \partial_\mu\xi_L - iV_\mu\xi_L + i\xi_L{\cal{L}}_\mu\,,
\nonumber\\
&&
D_\mu \xi_R
 = \partial_\mu\xi_R - iV_\mu\xi_R + i\xi_R{\cal{R}}_\mu\,,
\end{eqnarray}
with ${\cal{L}}_\mu$ and ${\cal{R}}_\mu$ being the external
gauge fields introduced by gauging $G_{\rm{global}}$.
The Lagrangian with the lowest derivatives is given by~\cite{hls}
\begin{equation}
{\mathcal L}_M
= F_\pi^2\mbox{tr}\left[ \hat{\alpha}_{\perp\mu}
  \hat{\alpha}_{\perp}^{\mu} \right]
{}+ F_\sigma^2\mbox{tr}\left[ \hat{\alpha}_{\parallel\mu}
  \hat{\alpha}_{\parallel}^{\mu} \right]
{}- \frac{1}{2g^2}\mbox{tr}\left[ V_{\mu\nu}V^{\mu\nu} \right]\,,
\label{lagmeson}
\end{equation}
where $g$ is the HLS gauge coupling
and the field strengths are defined by
$V_{\mu\nu} = \partial_\mu V_\nu - \partial_\nu V_\mu
{}- i\left[ V_\mu, V_\nu \right]\,$.
Expanding $\hat{\alpha}_{\perp \mu}$ and $\hat{\alpha}_{\parallel\mu}$,
\be
\hat{\alpha}_{\perp}^{\mu}
& = & \frac{1}{F_{\pi}}\partial_{\mu}\pi + {\mathcal A}^\mu
{}- \frac{1}{F_\pi}\left[ {\mathcal V}^\mu, \pi \right]
{}- \frac{1}{6F_{\pi}^{3}}\left[\left[\partial_{\mu}\pi, \pi\right],\pi\right]
{}+ \cdots,
\\
\hat{\alpha}_{\parallel}^{\mu}
& = &  \frac{1}{F_{\sigma}}\partial_{\mu} \sigma +{\mathcal V}^\mu - V^\mu
{}- \frac{i}{2F_{\pi}^{2}}\left[ \partial_{\mu} \pi, \pi \right]
{}- \frac{i}{F_\pi}\left[ {\mathcal A}^\mu, \pi \right]
{}+ \cdots\,,
\ee
where ${\mathcal V}^\mu = ({\mathcal R}^\mu + {\mathcal L}^\mu)/2$
and ${\mathcal A}^\mu = ({\mathcal R}^\mu - {\mathcal L}^\mu)/2$,
one finds the vector meson mass and the $\rho\pi\pi$ coupling constant as
\begin{eqnarray}
&&
m_V^2 = ag^2 F_\pi^2\,,
\quad
a = \frac{F_\sigma^2}{F_\pi^2}\,,
\\
&&
g_{\rho\pi\pi} = \frac{1}{2}ag\,.
\end{eqnarray}

The Lagrangian of mirror nucleons in the non-linear realization without
vector mesons was considered in~\cite{nemoto}. Its HLS-extended form
is found to be~\cite{Sasaki}
\be
\mathcal{L}_{N}
& = & \bar{Q}i\gamma^{\mu}D_{\mu}Q - g_{1}F_{\pi}\bar{Q}Q
{}+ g_{2}F_{\pi}\bar{Q}\rho_{3}Q
\nonumber\\
&& - im_{0}\bar{Q}\rho_{2}\gamma_{5}Q
{}+ g_{V} \bar{Q}\gamma^{\mu}\hat{\alpha}_{\parallel \mu}Q
{}+ g_{A}\bar{Q} \rho_{3} \gamma^{\mu}\hat{\alpha}_{\perp \mu}
\gamma_{5} Q\,,
\label{NLargrangian}
\ee
where the nucleon doublet
$Q = \left(\begin{array}{cc} Q_{1} \\ Q_{2} \end{array}\right)$ transforms as
\begin{equation}
Q \to h\,Q\,,
\end{equation}
the covariant derivative $D_\mu = \partial_\mu - iV_\mu$,
the $\rho_{i}$ are the Pauli matrices acting on the parity-doublet
and  $g_A$ and $g_V$ are dimensionless parameters.
To diagonalize the mass term in Eq.~(\ref{NLargrangian}), we transform $Q$ into
a new field $N$:
\be
\le N_{+} \\ N_{-} \re
= \frac{1}{\sqrt{2 \cosh \delta}} \le e^{\delta/2} & \gamma_{5}e^{-\delta/2}
\\ \gamma_{5}e^{-\delta/2} & -e^{\delta/2} \re
\le Q_{1} \\ Q_{2} \re\,,
\ee
where $\sinh \delta = - \frac{g_{1}F_{\pi}}{m_{0}}$.
We identify $N_{\pm}$ as parity even and odd states respectively.
The nucleon masses are found to be
\be
&& m_{N_{\pm}}
= \mp  g_{2} F_{\pi} + \sqrt{\left( g_{1}F_{\pi}\right)^{2} + m_{0}^{2}}\,,
\label{mass}
\\
&& \cosh \delta = \frac{\mn + \mm }{2m_{0}}\,.
\ee
Finally, we arrive at the Lagrangian in the parity eigenstate as
\begin{eqnarray}
\mathcal{L}_{N}
&=& \bar{N} i \sbar{D} N - \bar{N}  \hat{\mathcal{M}} N
{}+ g_{V}\bar{N}\gamma^{\mu} \hat{\alpha}_{\parallel\mu} N
{}+ g_A\bar{N}\gamma^\mu\hat{G}\hat{\alpha}_{\perp\mu}\gamma_5 N\,,
\label{Nlagrangian}
\\
\hat{\mathcal{M}}
&=& \left( \begin{array}{cc} \mn & 0 \\
0 & \mm \end{array} \right)\,,
\quad
\hat{G} = \left( \begin{array}{cc}
\tanh\delta & \gamma_5/\cosh\delta \\
\gamma_5/\cosh\delta & -\tanh\delta
\end{array}\right)\,.
\end{eqnarray}
The axial couplings for nucleons are
\be
g_{AN_{+}N_{+}} = - g_{AN_{-}N_{-}} = g_{A} \tanh \delta.
\ee

\setcounter{equation}{0}
\section{Chiral Invariant Mass of the Nucleon}
\label{sec:m0}

In this section, we calculate the decay width of $N_- \to \pi N_+$ to
extract the mass parameter $m_0$ and calculate the
pion-nucleon scattering lengths to be compared with the experimental data.

\subsection{Decay width of {\boldmath $N_{-} \rightarrow N_{+} + \pi$}}

The decay width at tree is given by
\be
\Gamma
&=& C_{2} \left( r \right)
\frac{\left( \mm -\mn \right)^{2} }{8 \pi F_{\pi}^{2}}
\sqrt{\left( \mm - \mn \right)^{2} - \mpi^{2}}
\left( \frac{g_{A}}{ \cosh \delta}\right)^{2}
\frac{\left[ \left( \mm + \mn \right)^{2} -\mpi^{2}
\right]^{\frac{3}{2}}}{2\mm^{3}}\,,
\nonumber\\
\label{gamma}
\ee
with $C_{2} \left( r \right) = T^{a}T_{a} = \frac{3}{4}$ in SU(2).
The nucleon axial-coupling in a linear sigma model for the parity doublers and
pions, given by $g_{AN_+N_+} = g_A\tanh\delta$, is typically smaller than unity.
In order to recover the well established experimental value, $1.267$, one should
add more states, in particular the $\Delta(1232)$ isobar, so that the
Adler-Weisberger sum rule is saturated by those resonances~\cite{DeTar}.

\begin{figure}
\centering
\includegraphics[angle=0, width=0.6\textwidth]{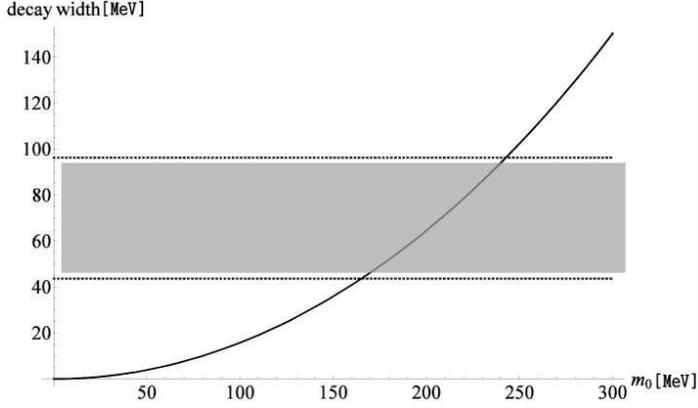}
\caption[0]{Decay width of $N_- \to N_+\pi$.
The gray colored region describes the experimental bound~\cite{pdg}.
\label{figure1}}
\end{figure}

In a non-linear sigma model, on the other hand, $g_A$ can be an
arbitrary parameter in Eq.~(\ref{NLargrangian}) to be determined from
the constraint,
\be
g_{AN_{+}N_{+}} = g_{A} \tanh \delta = 1.267\,.
\ee
Using the experimental values of $F_\pi = 92.42$ MeV, $m_{N_+} = 939$ MeV,
$m_{N_-} = 1535$ MeV,
one obtains the decay width as a function of $m_0$\footnote{Since $\left(\frac{g_{A}}{\cosh \delta} \right)^{2} = g_{A}^{2} \left( \tanh \delta \right)^{2}\left( \frac{1}{1- \frac{4 m_{0}^{2} }{\left( \mn + \mm \right)^{2}}} - 1 \right)$, setting $g_{A} \tanh \delta=1.267$, one finds that only '$m_{0}$' remains as the undetermined parameter.} as in Fig.~\ref{figure1}.
Using the decay width~\cite{pdg},
\be
\Gamma^{\rm exp}_{N_{-} \rightarrow N_{+} \pi} = 70 \pm 26.25\ \MeV\,,
\ee we get a bound of $m_{0}$ as
\be
m_{0} = 204 \pm 39\ \MeV\,.
\label{m0}
\ee
This corresponds to
\be
0.981 \lsim \tanh \delta \lsim 0.991\,.
\ee

\subsection{{\boldmath $\pi N $} scattering lengths}

\begin{figure}
\centering
\includegraphics[angle=0, width=0.8\textwidth]{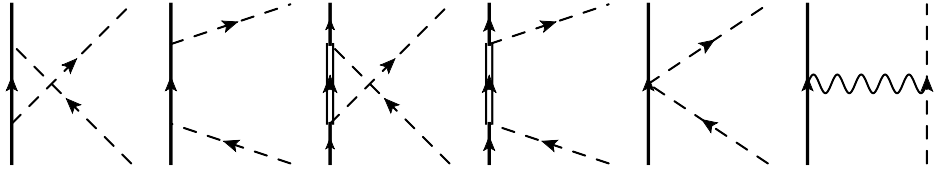}
\caption[0]{Diagrams relevant to $\pi N$ scattering lengths at the tree level.
The solid, dashed, double and wiggled lines describe nucleon, pion, parity
partner of the nucleon and $\rho$ meson respectively. }
\label{tree}
\end{figure}

The pion-nucleon scattering amplitude, $\pi^{a}(q) + N(p)
\rightarrow \pi^{b}(q^{\prime}) + N(p^{\prime})$, is written as
\be T^{ab} (p,q;p^{\prime}, q^{\prime}) = \bar{u}(p^{\prime})
\left[\left( A^{(+)} + \frac{1}{2} \left( \sbar{q}+
\sbar{q}^{\prime} \right) B^{(+)} \right) \delta_{ab}
{}+ \left( A^{(-)} + \frac{1}{2} \left( \sbar{q}+
\sbar{q}^{\prime} \right) B^{(-)} \right) i
\epsilon_{bac}\tau_{c}\right] u(p)\,,\nonumber \ee where $a$
and $b$ are isospin indices. The s-wave isospin-even and
isospin-odd scattering lengths are defined in terms of the
Mandelstam variables as \be a^{(\pm)}_{0} = \frac{1}{4 \pi
\left( 1+ m_{\pi}/ m_{N_{+}} \right)} \left( A^{(\pm)}_{0} +
m_{\pi} B^{(\pm)}_{0} \right)\,, \ee where the subscript $0$
indicates that we take $s = (\mn + \mpi )^{2}$, $t = 0$ and $u
= (\mn - \mpi)^{2}$. The tree diagrams contributing to the
scattering lengths are shown in Fig.~\ref{tree}.
Using the Feynman rules obtained from the Lagrangians (\ref{lagmeson}) and
(\ref{Nlagrangian}), one arrives at the following expressions:
\be
a_{0}^{+}
& = & \frac{1}{4\pi \left(1 + \frac{m_{\pi}}{m_{N_{+}}}\right)}
\left\{ -\left(\frac{g_{A}}{F_{\pi}} \tanh\delta \right)^{2}
\frac{ m_{N_{+}}m^{2}_{\pi} }{4m^{2}_{N_{+}}-m^{2}_{\pi}} \right.
\nonumber \\
&& \left.
{}+ \left( \frac{g_{A}}{F_{\pi}}\frac{1}{\cosh\delta} \right)^{2}
\frac{m_{\pi}^{2}\left(m_{N_{-}}-m_{N_{+}}\right)}
{2 \left( \left( m_{N_{-}}- m_{N_{+}} \right)^{2} - m_{\pi}^{2}\right)}
\right\}\,,
\label{evens}\\
a_{0}^{-}
& = & \frac{1}{4\pi \left( 1+\frac{m_{\pi}}{m_{N_{+}}} \right)}
\left\{ \left( \frac{g_{A}}{F_{\pi}} \tanh\delta \right)^{2}
\frac{m_{\pi}^{3}}{2\left( 4m_{N_{+}}^{2}-m_{\pi}^{2} \right)} \right.
\nonumber \\
&&
{}+ \left( \frac{g_{A}}{F_{\pi}} \frac{1}{\cosh\delta }  \right)^{2}
\frac{m^{3}_{\pi}}{2 \left[ \left( m_{N_{-}} - m_{N_{+}} \right)^{2}
{}- m_{\pi}^{2} \right]}
\nonumber \\
&& \left. {}+ g_{V}\frac{m_{\pi}}{2 F_{\pi}^{2}} +
\left(1-g_{V} \right) \frac{\mpi a g^{2}}{2 m_{\rho}^{2}}
\right\}\,. \label{odds} \ee We note that in the last line of
Eq.~(\ref{odds}) the terms with $g_V$ are precisely canceled
since $m_{\rho}^{2} = a g^{2} F_{\pi}^{2}$. This means that the
contribution from NN$\pi \pi$ interaction is canceled and
$\rho$ meson interaction gives the main contribution to the odd
scattering length.
This cancellation arises independently of any HLS parameters and therefore
it is purely the consequence of low-energy theorems of chiral symmetry.
Since the nonlinear sigma model (NL$\sigma$M) is gauge-equivalent to HLS,
all low-energy theorems of the NL$\sigma$M are encoded in the HLS
Lagrangian~\cite{HLStree}. The baryon HLS Lagrangian constructed here has
the HLS and hence has the same low-energy theorems of the baryonic NL$\sigma$M.

Using Eq.~(\ref{m0}) for $m_0$, we obtain the scattering lengths
\be
&& a_{0}^{+} = (-6.0 \pm 0.4) \times 10^{-5} \MeV^{-1}\,,
\\
&& a_{0}^{-} = ( 57 \pm 1) \times 10^{-5} \MeV^{-1}\,, \ee
which are to be compared the experimental
values\cite{Beane:2002wk},
\begin{equation}
a_{0,{\rm exp}}^+ = (- 2.446 \pm 0.504) \times 10^{-5} \MeV^{-1}\,,
\quad
a_{0,{\rm exp}}^- = ( 66.043 \pm 0.935) \times 10^{-5} \MeV^{-1}\,.
\end{equation}
Considering that these are tree-order results, they are not
unreasonable. The discrepancy from the experimental data could
be removed if other resonances, such as, e.g., the axial-vector
mesons are included as in ~\cite{gallas}.

\subsection{{\boldmath $m_{0}$} in various models}

In this section, we compare the values of $m_0$ evaluated in
the various approaches summarized in Table~\ref{tablem0}.

\begin{table}
\begin{center}
\begin{tabular}{|c|l|l|l|}
\hline
{} & $\bf{m_{0}}$ [MeV] \hspace*{0.5cm}
& \textbf{$\rho$(770)}\hspace*{0.7cm} & \textbf{$a_{1}$(1260)}
\\
\hline
\textbf{LSM (tree)}~\cite{DeTar}  & 270 & no & no
\\
\hline
\textbf{LSM (tree)}~\cite{gallas}  & 460 $\pm$ 136 & yes & yes
\\
\hline
\textbf{HBChPT (1-loop)}~\cite{nemoto} & 335 $\pm$ 95 & no  & no
\\
\hline
\textbf{this work (tree)}  & 204 $\pm$ 39 & yes & no
\\
\hline
\end{tabular}
\caption{The range of $m_{0}$ extracted in various approaches
in medium-free space. ``LSM'' stands for linear sigma model and
``HBChPT'' for heavy baryon chiral perturbation theory, respectively.
} \label{tablem0}
\end{center}
\end{table}

\begin{itemize}
\item {{\boldmath $m_0$}\ \bf in vacuum}

The chiral invariant mass, $m_0$, can be determined solely
from the decay width of $N(1535)$ to $\pi$ and $N(940)$ in
a linear sigma model without axial-vector mesons. For the
experimental value for the width $\Gamma \simeq 70$ MeV,
the model predicts $m_0 \simeq 270$
MeV~\cite{DeTar,gallas}. However the pion-nucleon
scattering lengths and the axial charges calculated in the
model do not come out in agreement with the existing data.
In particular, the isospin-even scattering length depends
strongly on the sigma-meson mass that cannot be pinned down
accurately.

In order to have a better fit, one introduces the
axial-vector meson $a_1$ in such a way that it couples to
the positive-parity nucleon differently from the
negative-parity state, as well as an independent coupling
strength of the vector (axial-vector) mesons to the
nucleons~\cite{gallas}. The free parameters can be varied
to reproduce the experimental data $\Gamma_{N_{-}
\rightarrow N_{+} \pi} = 67.5 \pm 23.6$ MeV, $\Gamma_{a_{1}
\rightarrow \pi \gamma} = 0.640 \pm 0.246$ MeV and
$g_{AN_{+}N_{+}} = 1.267 \pm 0.004$, and the lattice
measurement, $g_{AN_{-}N_{-}}^{\rm lattice} = 0.2 \pm
0.3$~\cite{takahashi}\footnote{It is perhaps unsafe to rely
on the sole measurement available for this quantity if the
prediction is sensitive to its value. Further lattice
measurements are needed.}. The model also gives the $\pi$-N
scattering lengths in reasonable agreement with
experiments~\cite{gallas}.

Using a three-flavor heavy-baryon chiral Lagrangian~\cite{hbchpt},  one-loop chiral perturbation calculation for widths was carried out for various channels~\cite{nemoto}. Within the given experimental uncertainties that reflect on the parameters of the Lagrangian, the $m_0$ extracted is found to range $240$ MeV $\lsim$ $m_{0}$ $\lsim$ $430$ MeV.
Here the error  comes mainly from a large uncertainty of the decay
width of $N^* \to \pi N$. The corresponding range of $\tanh\delta$ in
the width increases due to the loop corrections and thus $m_0$ becomes
slightly smaller than its tree-level value.

\item {{\boldmath $m_0$}\ \bf  in hot and dense matter:}

The linear sigma model with parity doublers has been applied to symmetric~\cite{LSMsym1,pisarski,LSMsym2} and asymmetric nuclear matter~\cite{LSMasym} in hot and dense environment within mean field approximation.
The model is constructed in such a way that at $T=0$ the properties of
nuclear matter, i.e. saturation, binding energy and incompressibility,
are correctly reproduced while satisfying low-energy theorems of chiral
symmetry. In order to have a reasonable value for incompressibility,
a large $m_0 \sim 800$ MeV is found to be required. This is substantially
changed when a scalar tetraquark-state is introduced, which makes the
incompressibility in an acceptable range with $m_0 \sim 500$ MeV~\cite{gallas2011}.

A different approach to determine $m_0$ has recently been proposed using
a chiral Lagrangian implementing conformal invariance, where the origin
of $m_0$ is mostly of the gluon condensate. With the chiral symmetry
restoration temperature taken to be $T_{\chi{\rm SR}} \sim 170$ MeV, at zero density,
the $m_0$ comes out to be $\sim 210$ MeV~\cite{Sasaki}. This coincides with
the result of the HLS model at tree obtained in this paper, Eq.~(\ref{m0}).

As stressed in Introduction, both the dilaton scalar and the vector mesons in dHLS are the relevant degrees of freedom in nuclear matter and could lead in the mean-field approximation to a better treatment of both nuclear matter and matter at higher density, thereby giving a better constraint on $m_0$ than so far arrived at, i.e., $\sim 500$-$800$ MeV. This work is presently being done and will be a subject of future publication.

\end{itemize}


\setcounter{equation}{0}
\section{Analysis of Renormalization Group Equations}
\label{sec:rge}

The phase structure of the HLS model has been studied based on
the renormalization group equations (RGEs) where loop effects
are systematically calculated using the chiral perturbation
theory including vector mesons at one-loop order~\cite{hls}. HLS renders a systematic chiral perturbation feasible in the presence of the vector
mesons. The key point is that the HLS coupling constant can be
taken as
\begin{equation}
g \sim {\mathcal O}(p)\,,
\end{equation}
which means that the vector meson mass is of ${\mathcal O}(p)$,
in the same chiral order as the pion mass. With the vector
meson mass $\sim 6$ times the pion mass in matter-free space,
this counting may appear to be unreasonable but it is justified
by that it goes as ${\mathcal O}(N_c^0)$ with the corrections
coming at ${\mathcal O}(1/N_c)$ and is endowed with the vector
manifestation fixed point of HLS at which the vector meson mass
goes to zero in the chiral limit. In fact it has been shown in
\cite{hls} that chiral perturbation expansion works even in
matter-free space as well as does nonlinear sigma model at
one-loop order to which the HLS expansion has been done. We
will take this as an indication that the extrapolation from
large $N_c$ to $N_c=3$ is a reliable approximation. In this
section we further extend it to a system with parity-doubled
nucleons. We will follow closely the quantization procedure of
Ref.~\cite{hls}. Details of the diagrammatic calculations and
the RGEs are relegated to Appendices~\ref{app:naive} and~\ref{app:loop}.
In this section we analyze the RGEs and their fixed-point
structure of the hidden local symmetric parity-doublet model.

In order to gain insight, it is helpful to examine the standard (or ``naive") assignment with $m_0=0$. The details are given in Appendix \ref{app:naive}. Here we summarize the key results of the analysis.

The four coupled equations~(\ref{md}), (\ref{ms}), (\ref{rgev:naive})
and (\ref{rgea:naive}) describe
the RGE flows of the masses $m_{N_\pm}$ and coupling constants
$g_A$ and $(1-g_V)$. It is a complicated set of equations and
has not yet been fully analyzed. There may be several fixed points or fixed lines. There is however one
strikingly simple fixed point which is easy to identify and is argued to be relevant to QCD and
that is what is called ``dilaton-limit fixed point (DLFP for
short)'' corresponding to the dilaton limit discussed in \cite{Sasaki}:
\be
\left( 1- g_{V}, \,g_{A} - g_{V}, M_S,
M_D \right) = \left( 0, 0, 0, 0 \right)\,
\ee
where $M_S=\frac 14 (m_{N_+}+m_{N_-})^2$ and $M_D=\frac 14 (m_{N_+}-m_{N_-})^2$.
This ensures that the vector mesons decouple from the nucleons
toward the dilaton limit encoded by $g_A=g_V=1$ and the suppression of the
repulsive force due to the vector meson exchange~\cite{Sasaki} remains
a robust statement at the quantum level. In fact, the DLFP is an infrared
fixed point, i.e.,
\\
\be \frac{\partial}{\partial M_{D}}
\left[ \mu\frac{d M_D}{ d \mu} \right]
& = &
6 \left(
\frac{1}{4 \pi F_{\pi}}
\right)^{2} \mu^{2} > 0\,,
\\
\frac{\partial}{\partial M_{S}} \left[ \mu\frac{d M_S}{ d \mu}
\right]
& = &
6 \left( \frac{1}{4 \pi F_{\pi}}
\right)^{2} \mu^{2} > 0\,,
\\
\frac{\partial}{\partial \left( 1- g_{V}
\right)}\left[ \mu \frac{d (1-g_V)}{d\mu}
\right]
& = &
\frac{ 1 }{\left( 4 \pi
F_{\sigma} \right)^{2}}
\left[ \left( 3 - a^2 +2a \right) \mu^{2}
{}+ m_{\rho}^{2} \right] > 0\,,
\\
\frac{\partial}{\partial g_A}
\left[ \mu\frac{d g_A}{ d \mu} \right]
&=&
\frac{4}{(4\pi F_\pi)^2}\mu^2 > 0\,,
\label{ddgv}
\ee
for which we set $
\left( 1- g_{V}, g_A-g_V,  M_S,
M_D\right)$ = $\left(
0, 0, 0, 0 \right)$.
A special value $a=2$ explains the vacuum phenomenology, such as the
vector meson dominance, although it is not a fixed point of the
RGE. The HLS theory possesses $a=1$ as a fixed point matching with
QCD taking $\langle \bar{q}q \rangle \to 0$~\cite{hls} and this is
not affected by the nucleons when $M_D=M_S=0$.
Thus, Eq.~(\ref{ddgv}) is always positive for any $a$ in
the range between $1$ and $2$.

Now turning to the case of the mirror assignment with $m_0\neq 0$, the situation
is a bit more involved since the treatment will depend upon whether $m_0$ is light
or heavy: Consider two extreme limits: (1) $m_0 \sim {\cal O}(m_\pi)$ and (2) $m_0 \gg
\Lambda_{\rm QCD}$.
In the case (1), we can treat $m_0$ as the small quantity as one does in ChPT and
ignore terms of ${\cal O}(m_0)$ and then the above analysis will apply.
In the case of (2), we apply heavy-baryon formalism as one does in heavy-baryon
chiral perturbation theory (HBChPT). As in HBChPT, ${\cal O} (m_0^{-1})$ is ignored.
For details, see Appendix B.
It is verified that the coupled equations do possess the infrared fixed point
and the dilaton-limit fixed point is intact for both cases of $m_0\approx 0$
and $m_0 \gg \Lambda_{\rm QCD}$,
\be
(1-g_V, g_V-g_A, M_S, M_D)=(0, 0, m_0^2, 0)\,,
\ee
as in the standard assignment.

The suppressed repulsive interaction associated with an IR fixed
point is therefore a common feature in the two different assignments,
``naive'' and mirror, of chirality. It is natural that the short-distance
interaction is independent of the chirality assignment. The physics behind it
must be related with some new symmetries which may dynamically emerge in hot/dense
matter. We consider that this is a manifestation of ``emergent symmetry" akin to
that associated with the Harada-Yamawaki vector manifestation in hidden local
symmetry~\cite{hls}. As shown in~\cite{Sasaki}, at the dilaton limit the
lowest-lying mesons (scalar, pseudo-scalar, vector and axial-vector) are
assembled into a full representation of chiral group. Weinberg's mended
symmetry~\cite{weinberg} -- which is not present in the fundamental QCD Lagrangian
but can be emergent due to in-medium collective excitations -- becomes manifest
there and this might protect the dilaton limit at quantum level. We note that
this aspect of emergence of symmetries, both for the dilaton-limit fixed point
and the vector manifestation fixed point, is absent in the approaches found in
the literature that are not anchored on hidden flavor gauge symmetry and represents
a falsifiable bona-fide prediction of HLS in dense and hot matter.

\setcounter{equation}{0}
\section{Remarks and Conclusions}
\label{sec:conclusions}
In this paper, we constructed a parity-doublet nonlinear sigma
model with hidden local symmetry (PDHLS model) and extracted at
the lowest order in chiral perturbation theory the
chiral-invariant mass $m_0$ from experimental data in the
vacuum. We found it to be $\sim 200$ MeV. By itself, this value
has little significance, since in nonlinear realization, the
chiral-invariant mass $m_0$ gets compounded into the physical
mass with important dynamically generated mass. Furthermore,
quantum loop corrections may not be ignored in the analysis.
However it turns out to play an important role in dense
baryonic matter as discussed in \cite{Sasaki} and re-stressed
below.

In calculating quantum loop corrections in the PDHLS model (the
detailed discussion of the results of which will be relegated to a future publication), we have discovered that the one-loop renormalization group
equations (RGE) have a fixed point that has not been so far observed in other approaches. The set of four coupled equations for the RGE flow of the nucleon masses $m_{N_\pm}$, the vector coupling $g_V$ and the axial-vector
coupling $g_A$ have a simple IR fixed point $(1-g_V, g_A-g_V, M_S, M_D)=(0, 0, m_0^2, 0)$ in the standard (with $m_0=0$) or mirror (with $m_0\neq 0$) assignment. It is very possible that these set of equations possess a variety of other fixed points, some of which may be consistent with QCD.
Our proposal here is that this fixed point that we refer to as
dilaton-limit fixed point (DLFP) is precisely the ``dilaton
limit" considered in \cite{Sasaki} that encodes the phenomenon of mended symmetries~\cite{weinberg}.

As discussed in \cite{Sasaki}, one way of driving the baryonic
system at zero temperature and low density ($n\lsim n_0$)
described by the HLS (or PDHLS) Lagrangian to a dense baryonic
system ($n\gg n_0$) is to introduce a scalar degree of freedom
in HLS (or PDHLS) in terms of the ``soft" dilaton associated
with the trace anomaly of QCD and then take the dilaton
limit~\cite{beane} to go over to the linearly realized
(Gell-Mann-L\'evy-type) sigma model Lagrangian. This dilaton
limit is found to correspond exactly to the DLFP we found
in the RGEs described above.~\footnote{Although a heuristic
consideration indicates that the fixed point is robust, we have
not however verified in detail that this fixed point remains
unaffected when the dilaton field $\chi$ is introduced into the
loop diagrams. The basic problem here is that scalar fields of
the $\chi$ type are problematic in higher order calculations.}
We interpret this exact correspondence as implying that as
density increases toward that of chiral restoration, the dense
matter flows toward this fixed point, the import of that point
being that the vector-meson--nucleon coupling gets suppressed at
high density. The major consequence of this fixed point is that
the strong hard-core repulsion present in nuclear interactions
(aptly described in terms of $\omega$ exchanges between two or
more nucleons) and the $\rho$ tensor force contributing crucially to the symmetry energy in neutron-rich systems as in compact stars will be strongly suppressed. How this prediction will affect the equation of state for compact stars is a very important issue to be worked out.

An intriguing possibility that can be entertained here is that
the DLFP and the vector
manifestation fixed point (VM) of HLS~\cite{hls} may be
intricately linked. In HLS, the fixed point $(a, g)=(1, 0)$
gets linked to QCD via the matching of QCD current correlators
at the matching scale by identifying the
$\la\bar{q}q\ra\rightarrow 0$ limit with the HLS coupling
$g\rightarrow 0$ limit. We conjecture that the DLFP linked to conformal symmetry is an IR fixed point that is reached {\em before} the VM/HLS is reached. This is quite analogous to the limit $a\rightarrow 1$ before reaching the ``vector limit" $g=0$ as Georgi discussed for vector symmetry
in HLS~\cite{georgi}.
Unlike in the case of the VM/HLS where the matching of
correlators enables one to make contact with QCD, here we are
not making any direct ``match" with QCD. We are assuming that
the dilaton limit imposed on effective field theory which works
well at the low density commensurate with nuclear matter is
consistent with the phase structure of QCD at high density
where chiral phase transition is to take place. As stated, we cannot rule out the possibility that the set of our RGEs have
other fixed points that are not inconsistent with QCD. This
issue will be an object of future study.

Now a comment on the role of a non-zero chiral invariant mass
$m_0$ in nuclear physics.  When nuclear matter is described in
effective field theory anchored on the low-momentum nuclear interaction
$V_{low-k}$ derived via
Wilsonian renormalization group equations, how the nucleon mass
scales in medium as a function of density turns out to be quite
important for the structure of finite nuclei as well as nuclear
matter~\cite{kuoetal,dklr}. The presence of a substantially big
$m_0$ as is found in some analysis in medium~\cite{pisarski}
would affect crucially how the nucleon mass scales in density
in such effective field theory approach to nuclei, nuclear
matter and dense compact-star matter.

\subsection*{Acknowledgments}

We acknowledge partial support by the WCU project
of the Korean Ministry of Educational Science and Technology
(R33-2008-000-10087-0). Part of this work was done when three of us (WGP, HKL and MR) were participating in the WCU-YITP Molecule Collaboration 18 April-18 May 2011. The work of C.S. has been partly supported by the Hessian LOEWE initiative
through the Helmholtz International Center for FAIR (HIC for FAIR).

\newpage

\appendix

\centerline{\large\bf  APPENDIX}

\setcounter{section}{0}
\renewcommand{\thesection}{\Alph{section}}
\setcounter{equation}{0}
\renewcommand{\theequation}{\Alph{section}.\arabic{equation}}

\section{Chiral Perturbation Theory in the Standard Assignment}
\label{app:naive}
Consider the standard (or ``naive") assignment with $m_0=0$. In the strict $m_0=0$ limit, the parity doublet decouple but here we take a non-zero $m_0$ and set it equal to zero at the end. This exercise allows us to gain some insight into what happens in the mirror scenario for the case when $m_0$ is small. We will consider the system to be near chiral restoration and assume that the entire mass of the nucleons is generated by spontaneous chiral symmetry breaking, with  $m_{N_\pm}$ vanishing at its restoration point.
We can then assign the chiral counting ${\mathcal O}(p)$ to the mass:
\begin{equation}
m_{N_\pm} \sim {\mathcal O}(p)\,.
\end{equation}
%
\begin{figure}
\centering
\includegraphics[angle=0, width=0.5\textwidth]{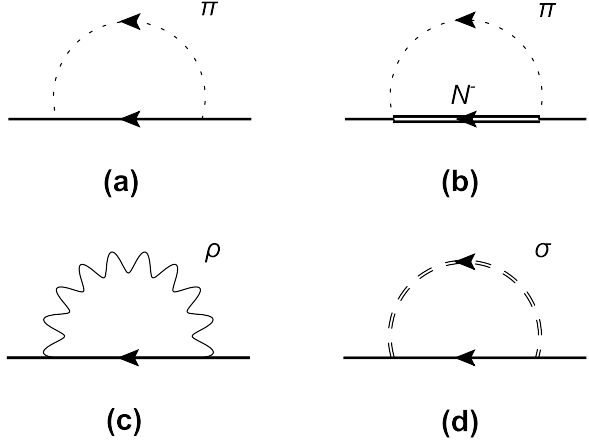}
\caption[0]{Diagrams of the nucleon self-energy: the thin
solid line corresponds to the nucleon and the double line to its odd-parity partner.
\label{figure4}}
\end{figure}
Evaluating the one-loop diagrams Fig.~\ref{figure4} in the relativistic formalism, one finds the RGEs of the nucleon masses as
\be
\mu\frac{d M_{D}}{ d \mu}
& = &
\frac{ 3 g_{A}^2 }{8\pi^2 F_{\pi}^2}
M_{D} \left[ \mu^{2} - M_{D} - 3 M_{S}
\right]
\nonumber \\
&&
{}- \frac{9\left(1 - g_{V} \right)^{2}}{16\pi^2} g^{2}
M_{D}\,,\label{md}
\ee
\be
\mu \frac{d M_{S}}{d \mu}
& = &
\frac{3g_{A}^2}{8\pi^2 F_{\pi}^2}
M_{S} \left( \mu^{2} - M_{S} -3 M_{D} \right)
\nonumber \\
&&
{}- \frac{9\left(1 - g_{V} \right)^{2}}{16\pi^2} g^{2}
M_{S}\,,\label{ms}
\ee
which indicate the dynamically generated masses vanish at the fixed point.
The RGEs of the Yukawa couplings from Figs.~\ref{VNN2}-\ref{ANN2}, in the $m_{0} \rightarrow 0$ limit, are obtained as
\begin{eqnarray}
\mu \frac{d}{d\mu} \left( 1- g_{V} \right)
& = &
\frac{m_{N_+}^2}{8\pi^2F_\pi^2}\tilde{\mathcal F}_0
{}+ \left( 1-g_V \right)\frac{1}{8\pi^2}\tilde{\mathcal F}_1
{}+ \left( g_V-g_A^2 \right)\frac{1}{8\pi^2}\tilde{\mathcal F}_2\,,
\label{rgev:naive}
\\
\mu \frac{dg_A}{d\mu}
& = &
\frac{m_{N_+}^2g_A}{8\pi^2F_\pi^2}\tilde{\mathcal G}_0
{}+ \left( 1-g_V \right)\frac{g_A}{8\pi^2}\tilde{\mathcal G}_1
{}+ \left( g_V-g_A^2 \right)\frac{g_A}{8\pi^2}\tilde{\mathcal G}_2\,,
\label{rgea:naive}
\end{eqnarray}
where $\mn$ becomes zero when $M_D=0$ and $M_{S}=0$. It is easy to see that $(1-g_V,g_A-g_V, M_S,M_D)=(0,0,0,0)$ is the fixed point of the coupled RGEs.
The explicit expressions of $\tilde{\mathcal F}_i$ and $\tilde{\mathcal G}_i$
are given by
\begin{eqnarray}
\tilde{\mathcal F}_0
&=&
\frac{1}{4}\left( a g_A^2 + \frac{g_V^2}{a} \right)\,,
\nonumber\\
\tilde{\mathcal F}_1
&=&
\left[ \frac{g_A^2}{F_\pi^2} + \frac{g_V\left(1+2g_V\right)}{2F_\sigma^2}
\right]\mu^2
{}- \frac{3}{2}\left( \frac{g_A^2}{F_\pi^2} + \frac{g_V^2}{F_\sigma^2}
\right)m_{N_+}^2
{}+ g^2\left( 2 - \frac{3}{2}g_V \right)\,,
\nonumber\\
\tilde{\mathcal F}_2
&=&
\frac{a}{2F_\pi^2}\mu^2\,,
\\
\tilde{\mathcal G}_0
&=&
\frac{1}{4}\left( g_A^2 + \frac{g_V^2}{a} + 2g_V\right)\,,
\nonumber\\
\tilde{\mathcal G}_1
&=&
\left( \frac{2}{F_\pi^2} + \frac{1-g_V}{F_\sigma^2}\right)\mu^2
{}+ \frac{2g_V}{F_\sigma^2}m_{N_+}^2
{}- \frac{5}{2}ag^2\,,
\nonumber\\
\tilde{\mathcal G}_2
&=&
-\frac{1}{F_\pi^2}\left( \mu^2 - 2m_{N_+}^2\right)\,.
\end{eqnarray}

%
\begin{figure}
\centering
\includegraphics[angle=0, width=0.5\textwidth]{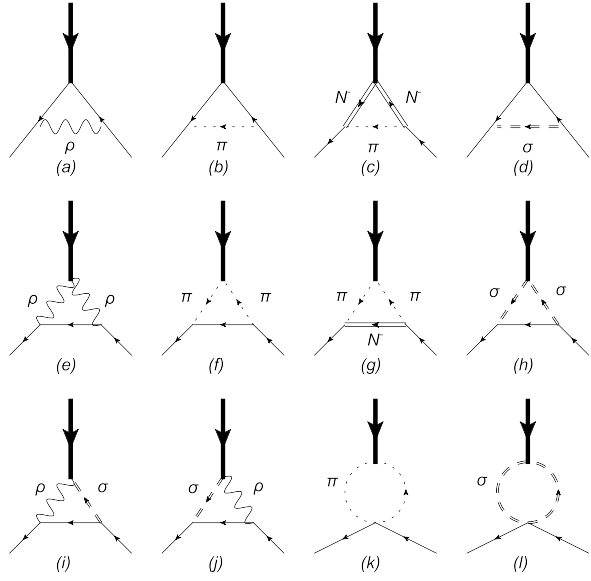}
\caption[0]{One-loop diagrams contributing to the $\bar{V}N_{+}N_{+}$ three-point
function.
\label{VNN2}}
\end{figure}
%
\begin{figure}
\centering
\includegraphics[angle=0, width=0.5\textwidth]{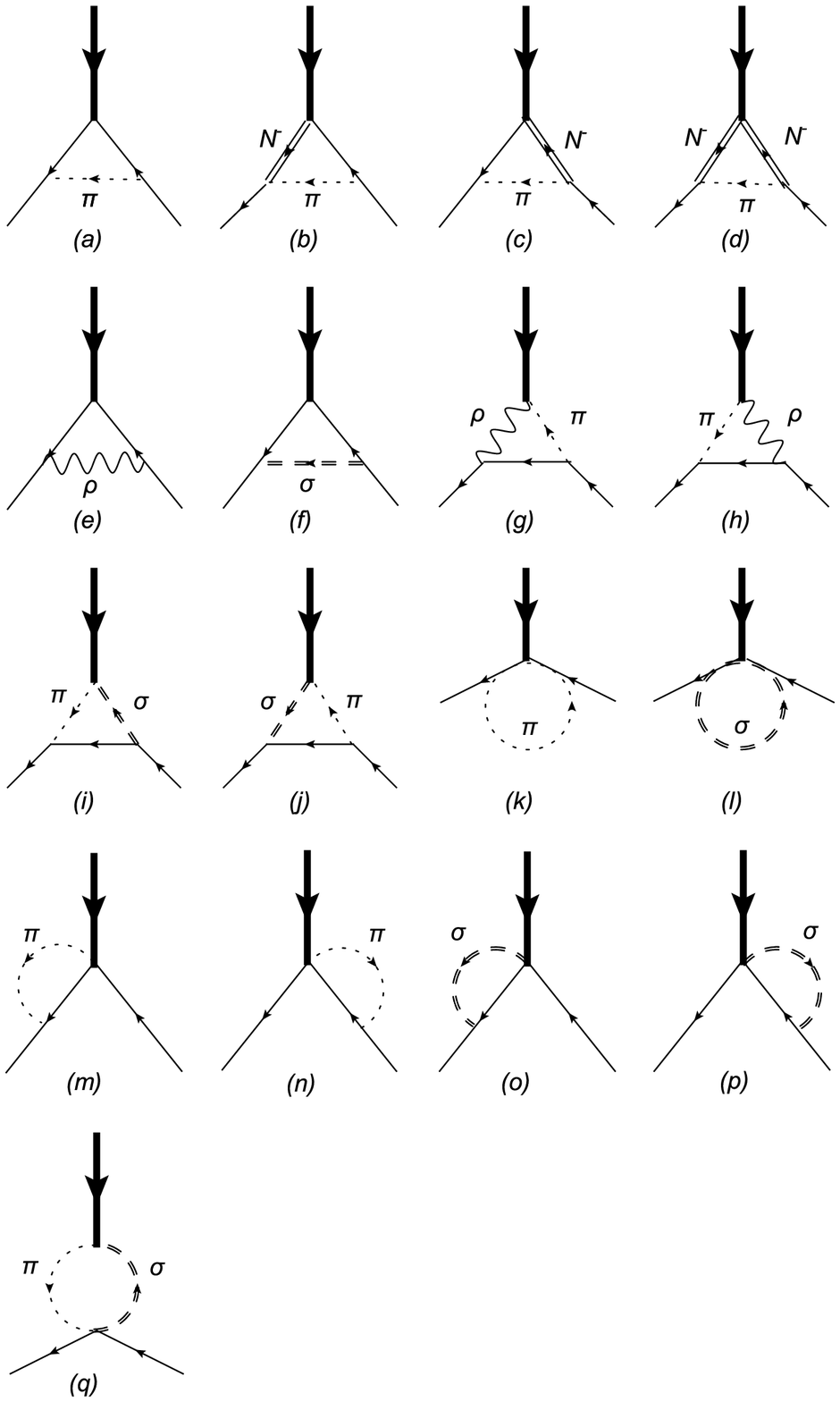}
\caption[0]{One-loop diagrams contributing to the $\bar{\mathcal{A}}N_{+} N_{+}$
vertex.
\label{ANN1}}
\end{figure}
%
\begin{figure}
\centering
\includegraphics[angle=0, width=0.5\textwidth]{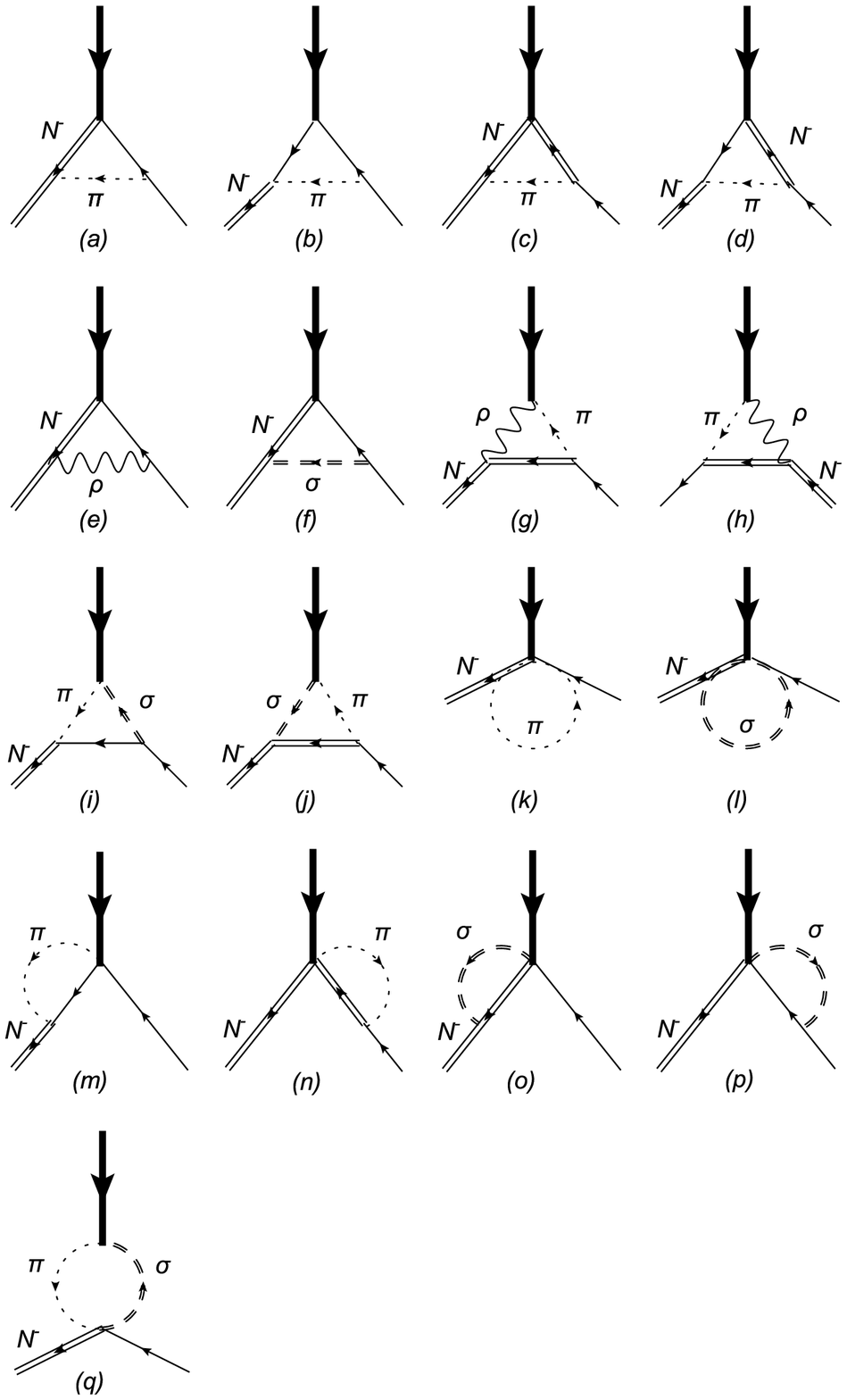}
\caption[0]{One-loop diagrams contributing to the $\bar{\mathcal{A}}N_{-} N_{+}$
vertex.
\label{ANN2}}
\end{figure}
%

The RGEs of $F_\pi^2, a$ and $g$ are obtaind as
\begin{eqnarray}
\mu\frac{dF_\pi^2}{d\mu}
&=&
\frac{1}{(4\pi)^2}\left[
3a^2 g^2 F_\pi^2 + 2(2-a)\mu^2
\right]
{}- \frac{g_A^2}{2\pi^2}
\left( m_{N_+}^2 + m_{N_-}^2 \right)\,,
\\
\mu\frac{da}{d\mu}
&=&
-\frac{1}{(4\pi)^2}(a-1)\left[
3a(1+a)g^2 - (3a-1)\frac{\mu^2}{F_\pi^2}
\right]
{}+ \frac{g_A^2}{2\pi^2}\frac{a}{F_\pi^2}
\left( m_{N_+}^2 + m_{N_-}^2 \right)\,,
\\
\mu\frac{dg^2}{d\mu}
&=&
-\frac{1}{(4\pi)^2}\frac{87-a^2}{6}g^4
{}+ \frac{1}{6\pi^2}\left( 1 - g_V \right)^2 g^4\,,
\end{eqnarray}
which agree with the expressions given in~\cite{hkr} when the nucleons
are replaced with constituent quarks.

\setcounter{equation}{0}
\section{Chiral Perturbation Theory in the Mirror Assignment}
\label{app:loop}

In the mirror assignment the nucleon mass is not entirely generated by
spontaneous chiral symmetry breaking and we identify the origin of
the chiral invariant mass $m_0$ with the explicit breaking of the QCD
scale invariance, i.e. a hard dilaton~\cite{Sasaki}, which has no direct link  with the chiral dynamics. Consider $m_0$ to be large compared with dynamically generated mass and adopt a heavy baryon chiral perturbation theory (HBChPT)~\cite{HBChPT} in the presence of
$m_0$. We write the nucleon momentum as
\begin{equation}
p^\mu = m_0 v^\mu + k^\mu\,,
\end{equation}
where $v^\mu$ is the four-velocity with $v^2=1$ and $k^\mu$ is the
residual momentum of order $\Lambda_{\rm QCD}$, so that one can
perform a chiral perturbation theory systematically in energy range
below the chiral symmetry breaking scale $\Lambda_{\chi} \sim 1$ GeV.
A heavy baryon field $B$ is defined by~\footnote{
This is a slightly different definition from that introduced in~\cite{nemoto}.
}
\begin{equation}
\begin{pmatrix}
B_+
\\
B_-
\end{pmatrix}
=
\exp\left[ i m_0 v\cdot x \right]
\begin{pmatrix}
N_+
\\
N_-
\end{pmatrix}\,.
\label{heavy}
\end{equation}
The Lagrangian (\ref{Nlagrangian}) is rewritten as~\footnote{
 In HBChPT in the mirror assignment, we treat $\frac{p}{m_{0}}$
 as a small quantity.
 We calculate the two- and three-point
 functions with the Largrangian (\ref{HBChPT}) and obtain the RGEs for
 $M_{S}$, $M_{D}$, $1-g_{V}$ and $g_{A}$. Then,
 we take the leading order terms of ${\cal O} \left(
 \frac{p^{2}}{\left( 4 \pi F_{\pi} \right)^{2}} \right)$ and drop the terms
 of ${\cal O}\left(\frac{p^{2}}{\left( 4 \pi F_{\pi} \right)^{2}} \,
 \left(\frac{p}{m_0}\right)^{n} \right)$ with n $ \geq $ 1 after expanding
 $\tanh^{2} \delta$, $\frac{1}{\cosh^{2} \delta}$ and $\Delta m_{\pm}^{2}$
 in $\frac{p}{m_0}$.
}
\begin{eqnarray}
{\mathcal L}_N
&=&
i\bar{B}v^\mu D_\mu B - \Delta m_+ \bar{B}_+ B_+
{}- \Delta m_- \bar{B}_- B_-
\nonumber\\
&&
{}+ g_V\bar{B}v^\mu\hat{\alpha}_{\parallel\mu}B
{}+ g_A\bar{B}\left( 2S^\mu\rho_3 \tanh\delta
{}+ v^\mu\rho_1 \frac{1}{\cosh\delta} \right)
\hat{\alpha}_{\perp\mu}B\,, \label{HBChPT}
\end{eqnarray}
where $S^\mu$ is the spin operator and
\begin{equation}
\Delta m_\pm = m_{N_\pm} - m_0\,.
\end{equation}
Note that because of the reduction~(\ref{heavy}) two small scales
comparable to $\Lambda_{\rm QCD}$, $\Delta m_\pm$, appear in the mass term.

 For our calculation in the background field gauge,
we again follow the notations of Harada and Yamawaki~\cite{hls} for quantum
and background fields. In case of without a negative parity nucleon,
further details can be found in, e.g.~\cite{PMR}.

\subsection{Quantum corrections to the Yukawa couplings}

In this subsection, we show the $\bar{V}B_{+}B_{+}$ three-point functions and give the regularization condition for $\left( 1 - g_{V} \right) $, where $\bar{V}_\mu$ is the background field of the HLS gauge field $V_\mu$.
The renormalized coupling is given by
\begin{equation}
(1-g_V)_{\rm bare} = Z_{3V}\left(Z_N Z_V^{1/2}\right)^{-1}(1-g_V)\,,
\end{equation}
where $Z_V$ represents the wavefuntion renormalization of the vector field
and $Z_{3V}$ appears in the counter term of the vector interaction.
Expanding $Z_{N,3V} = 1 + Z_{N,3V}^{(1)} \cdots$ and using $Z_V = 1$ for
the classical field $\bar{V}^\mu$, one obtains
the regularization condition as
\begin{equation}
(1-g_V)_{\rm bare}
{}+ (1-g_V)\left( - Z_{3V}^{(1)} + Z_N^{(1)} \right)
= \mbox{finite}\,.
\label{regular VNN}
\end{equation}
The counter terms are calculated from the 1-loop graphs of the two- and
three-point functions:
\begin{equation}
Z_{3V}^{(1)} = \Gamma_{\bar{V}B_+ B_+}|_{\rm div}/(1-g_V)\,,
\end{equation}
where the $\Gamma_{\bar{V} B_{+} B_{+}}$ is the sum of 1-loop diagrams given
in Fig.~\ref{VNN2}.

Other relevant three-point functions which give the loop corrections to the
axial coupling $g_A$ are the $\bar{{\mathcal A}}B_+ B_+$ and
$\bar{{\mathcal A}}B_+ B_-$ functions. The 1-loop graphs are shown
in Figures~\ref{ANN1} and \ref{ANN2}.

The renormalization conditions read
\be
(g_A\tanh\delta)_{\rm bare}
{}+ g_A\tanh\delta\left( - Z_{3A++}^{(1)} + Z_{N_{+}}^{(1)}
\right)
= \mbox{finite}\,,
\label{reg:tanh}
\ee
and
\be
\left( \frac{g_A}{\cosh\delta} \right)_{\rm bare}
{}+ \frac{g_A}{\cosh\delta}\left[ - Z_{3A+-}^{(1)}
{}+ \frac{1}{2}\left( Z_{N_{+}}^{(1)} + Z_{N_{-}}^{(1)} \right)
\right]
= \mbox{finite}\,,
\label{reg:cosh}
\ee
with
\begin{eqnarray}
Z_{3A++}^{(1)}
= \Gamma_{\bar{\mathcal A}B_+B_+}|_{\rm div}/(g_A\tanh\delta)\,,
\quad
Z_{3A+-}^{(1)}
= \Gamma_{\bar{\mathcal A}B_+B_-}|_{\rm div}/(g_A/\cosh\delta)\,.
\end{eqnarray}

\subsection{Renormalization group equations for the 2- and 3-point functions}
In the ordinary HBChPT without a negative parity nucleon, the leading
order parameters, $m_N$ and $g_A$, are not renormalized in the chiral limit.
In the present perturbation theory with the reduction~(\ref{heavy}),
$\Delta m_\pm$ appears as the small scales which remain non-vanishing
in the chiral limit. Therefore, the nucleon masses and coupling constants
receive loop corrections proportional to $\Delta m_\pm$. Using the standard
technique to evaluate the loop integrals~\cite{PMR}, one obtains the RGE
of $M_D$ as
\begin{equation}
\mu\frac{dM_D}{d\mu}
=
6\left(\frac{g_A}{4\pi F_\pi \cosh\delta}\right)^2
M_D\left( \mu^2 + 8M_D \right)\,, \label{rgemd}
\end{equation}
whereas $M_S$ does not evolve with the loop effect at this order, i.e.
\begin{equation}
\mu\frac{dM_S}{d\mu} = 0\,. \label{rgems}
\end{equation}

The $\bar{{\cal A}} B_{+} B_{+}$ and $\bar{{\cal A}} B_{+}
B_{-}$ functions computed from Figs.~\ref{ANN1} and \ref{ANN2} give the RGEs of $g_{A} \tanh \delta$ and $g_{A}$/$\cosh \delta$. Using the identity,
\be \mu
\frac{d}{d\mu} \left( \frac{ g_{A}^{2} }{ \cosh^{2} \delta}
\right) + \mu \frac{d}{d\mu}\left( g_{A}^{2} \tanh^{2} \delta
\right) = \mu \frac{d}{d\mu} \left( g_{A}^{2} \right)\,,
\ee
the RGE of $g_A$ is derived. {}From Eqs.(\ref{regular VNN}),
(\ref{reg:tanh}) and (\ref{reg:cosh}), one arrives at the RGEs
of $(1-g_{V})$ and $g_{A}$:
\begin{eqnarray}
\mu \frac{d}{d\mu} \left( 1-
g_{V} \right)
& = &
\frac{1}{8\pi^2}{\mathcal F}_0
{}+ \left( 1-g_V \right)\frac{1}{8\pi^2}{\mathcal F}_1
{}+ \left( g_V-g_A^2 \right)\frac{1}{8\pi^2}{\mathcal F}_2\,,
\label{rgev}
\\
\mu \frac{dg_A}{d\mu}
& = &
\frac{g_A}{8\pi^2}{\mathcal G}_0
{}+ \left( 1-g_V \right)\frac{g_A}{8\pi^2}{\mathcal G}_1
{}+ \left( g_V-g_A^2 \right)\frac{g_A}{8\pi^2}{\mathcal G}_2\,,
\label{rgea}
\end{eqnarray}
where ${\mathcal F}_i$ and ${\mathcal G}_i$
are functions of the
parameters, $F_\pi, a, g, g_A, g_V$ and $m_{N_\pm}$.
The functions ${\mathcal F}_0$ and ${\mathcal G}_0$ are given by
\begin{eqnarray}
{\mathcal F}_0
&=&
\frac{ag_A^2}{F_\pi^2}\left[
\left( \mu^2 + 3\Delta m_+^2\right)\tanh^2\delta
{}+ \frac{\Delta m_-^2}{\cosh^2\delta}
\right]
{}+ \frac{g_V^2}{F_\sigma^2}\Delta m_+^2\,,
\\
{\mathcal G}_0
&=&
\left( -\frac{g_A^2}{F_\pi^2\cosh^2\delta} + \frac{2g_V}{F_\pi^2} \right)\mu^2\tanh^2\delta + \frac{3 g_A^2}{4F_{\pi}^{2}\cosh^2\delta}
\left( \Delta m_+^2 + \Delta m_-^2 + \frac{2}{3}\Delta m_+ \Delta m_-\right)
\nonumber\\
&&
{}+ \frac{g_A^2}{F_\pi^2}\tanh^2\delta
\left[ 4\Delta m_+^2\tanh^2\delta - \frac{1}{\cosh^2\delta}
\left( \Delta m_+^2 - 2\Delta m_-^2 + \Delta m_+\Delta m_- \right)
\right]
\nonumber\\
&&
{}+ \frac{g_V^2}{F_\sigma^2}\left[
2\Delta m_+^2\tanh^2\delta + \frac{3}{4\cosh^2\delta}
\left( \Delta m_+^2 + \Delta m_-^2 + \frac{2}{3}\Delta m_+ \Delta m_-\right)
\right]
\nonumber\\
&&
{}- \frac{g_V}{F_\pi^2}\left[
2\Delta m_+^2\tanh^2\delta - \frac{1}{\cosh^2 \delta}
\left( \Delta m_+^2 + \Delta m_-^2 \right)
\right]\,,
\end{eqnarray}
and therefore vanish when $(M_S,M_D)=(m_0^2,0)$ whereas ${\mathcal F}_{1,2}$
and ${\mathcal G}_{1,2}$ remain non-vanishing in this choice of parameters.
Expanding $\tanh^{2} \delta$, $\frac{1}{\cosh^{2} \delta}$ and $\Delta m_{\pm}^{2}$
in $\frac{1}{m_0}$,
the RGEs, (\ref{rgemd}), (\ref{rgems}), (\ref{rgev}) and (\ref{rgea}), are given
in the leading order of $\frac{1}{m_{0}}$ by
\begin{eqnarray}
\mu \frac{d M_{D} }{d\mu }
& = &
\frac{3g_{A}^2}{8 \pi^2 F_{\pi}^2 } M_{D} \left( \mu^{2} + 8 M_{D} \right) \left[ 1  + {\cal O} \left( \frac{1}{m_{0}^{2}}\right) \right] \,,
\label{rgemp0}\\
\mu \frac{d M_{S} }{d \mu } & = & 0\,,
\label{rgemm0} \\
\mu \frac{d}{d\mu} \left( 1-
g_{V} \right)
& = &
\left[ \frac{M_{D}}{8\pi^{2} F_{\pi}^{2}} \bar{\mathcal F}_0  + \frac{\left( 1-g_V \right)}{8\pi^{2}} \bar{\mathcal F}_1
{}+ \frac{\left( g_V-g_A^2 \right)}{8\pi^{2}} \bar{\mathcal F}_2 \right] \left[ 1 + {\cal O} \left( \frac{1}{m_{0}}\right) \right]\,,
\label{rgev0}
\\
\mu \frac{dg_A}{d\mu}
& = &
\left[ \frac{M_{D} g_{A} }{8\pi^{2} F_{\pi}^{2}} \bar{\mathcal G}_0  + \left( 1-g_V \right) \frac{ g_{A}}{8\pi^{2}} \bar{\mathcal G}_1
{}+ \left( g_V-g_A^2 \right) \frac{ g_{A} }{8\pi^{2}} \bar{\mathcal G}_2 \right] \left[ 1 + {\cal O} \left( \frac{1}{m_{0}}\right) \right]\,,
\label{rgea0}
\nonumber\\
\end{eqnarray}
where $\bar{\mathcal F}_i$ and $\bar{\mathcal G}_i$ are given by
\begin{eqnarray}
\bar{\mathcal F}_0
&=&
ag_A^2 + \frac{g_V^2}{a}\,,
\nonumber\\
\bar{\mathcal F}_1
&=&
\left[ \frac{g_A^2}{F_\pi^2} + \frac{g_V\left(1+2g_V\right)}{2F_\sigma^2}
\right]\mu^2
{}+ 6 M_{D} \left( \frac{g_A^2}{F_\pi^2} + \frac{g_V^2}{F_\sigma^2} \right)
\nonumber\\
&&
{}- g^2\left( 4 - \frac{15}{2}g_V + 3g_V^2\right)\,,
\nonumber\\
\bar{\mathcal F}_2
&=&
\frac{a}{2F_\pi^2}\mu^2\,,
\\
\bar{\mathcal G}_0
&=&
g_{A}^{2} + \frac{g_{V}^{2}}{a}
{}+ 2 g_{V}\,,
\nonumber\\
\bar{\mathcal G}_1
&=&
\left( \frac{2}{F_\pi^2} + \frac{1-g_V}{F_\sigma^2}\right)\mu^2
{}- \frac{4g_V}{F_\sigma^2} M_{D}
\nonumber\\
&&
{}- g^2\left[ 3\left( 1 - g_V\right) + \frac{5}{2} a \right]\,,
\nonumber\\
\bar{\mathcal G}_2
&=&
-\frac{1}{F_\pi^2}\left(
\mu^2 + 4 M_{D}
\right)\,,
\end{eqnarray}
and all terms in the RGEs are ${\cal O} \left( \frac{p^{2}}{\left( 4 \pi F_{\pi} \right)^{2}} \right)$ in chiral counting.
With Eqs.~(\ref{rgemp0}), (\ref{rgemm0}), (\ref{rgev0}) and (\ref{rgea0}),
we arrive at the fixed point
$\left( 1-g_{V}, g_{A} - g_{V}, M_S, M_{D} \right) = \left( 0, 0, m_0^2, 0 \right)$ in the large $m_{0}$ limit.



\begin{thebibliography}{99}

\bibitem{HLStree}
  M.~Bando, T.~Kugo, S.~Uehara, K.~Yamawaki and T.~Yanagida,
  Phys.\ Rev.\ Lett.\  {\bf 54}, 1215 (1985),
  M.~Bando, T.~Kugo and K.~Yamawaki,
  Phys.\ Rept.\  {\bf 164}, 217 (1988).

\bibitem{hls}
  M.~Harada and K.~Yamawaki,
  Phys.\ Rept.\  {\bf 381}, 1 (2003).



\bi{Son} D.T. Son and M.A. Stephanov,
Phys. Rev.
{\bf D69}, 065020 (2004).

\bi{SS}  T.~Sakai and S.~Sugimoto,
  Prog.\ Theor.\ Phys.\  {\bf 113}, 843 (2005).
  Prog.\ Theor.\ Phys.\  {\bf 114}, 1083 (2005).

\bi{HMY} M.~Harada, S.~Matsuzaki and K.~Yamawaki,
  Phys.\ Rev.\  D {\bf 82}, 076010 (2010).

\bi{BR91}  G.~E.~Brown and M.~Rho,
  Phys.\ Rev.\ Lett.\  {\bf 66}, 2720 (1991).

\bi{hkr} M.~Harada, Y.~Kim and M.~Rho,
  Phys.\ Rev.\  D {\bf 66}, 016003 (2002).


\bi{LR} H.~K.~Lee and M.~Rho,
  Nucl.\ Phys.\  A {\bf 829}, 76 (2009).

\bi{beane} S.~R.~Beane and U.~van Kolck,
  Phys.\ Lett.\  B {\bf 328}, 137 (1994).

\bibitem{Sasaki}
  C.~Sasaki, H.~K.~Lee, W.~G.~Paeng and M.~Rho,
  Phys.\ Rev.\  D {\bf 84}, 034011 (2011).

\bibitem{DeTar}
  C.~E.~Detar and T.~Kunihiro,
  Phys.\ Rev.\  D {\bf 39}, 2805 (1989).

\bibitem{mirror}
  D.~Jido, Y.~Nemoto, M.~Oka and A.~Hosaka,
  Nucl.\ Phys.\  A {\bf 671}, 471 (2000),
  D.~Jido, T.~Hatsuda and T.~Kunihiro,
  Phys.\ Rev.\ Lett.\  {\bf 84}, 3252 (2000),
  D.~Jido, M.~Oka and A.~Hosaka,
  Prog.\ Theor.\ Phys.\  {\bf 106}, 873 (2001).


\bi{LPR} H.~K.~Lee, B.~Y.~Park and M.~Rho,
  Phys.\ Rev.\  C {\bf 83}, 025206 (2011).

\bi{BR tensor} G.~E.~Brown and M.~Rho,
  Phys.\ Lett.\  B {\bf 237}, 3 (1990).

\bibitem{nemoto}
  Y.~Nemoto, D.~Jido, M.~Oka and A.~Hosaka,
  Phys.\ Rev.\  D {\bf 57}, 4124 (1998).


\bibitem{pdg}
  K.~Nakamura {\it et al.}  [Particle Data Group],
  J.\ Phys.\ G {\bf 37}, 075021 (2010).

\bibitem{Beane:2002wk}
  S.~R.~Beane, V.~Bernard, E.~Epelbaum, U.~G.~Meissner and D.~R.~Phillips,
  Nucl.\ Phys.\  A {\bf 720}, 399 (2003).
  U.~G.~Meissner, U.~Raha and A.~Rusetsky,
  Phys.\ Lett.\  B {\bf 639}, 478 (2006).

\bibitem{gallas}
S.~Gallas, F.~Giacosa and D.~H.~Rischke,
  Phys.\ Rev.\  D {\bf 82}, 014004 (2010).


\bibitem{takahashi}
  T.~T.~Takahashi and T.~Kunihiro,
  Phys.\ Rev.\  D {\bf 78}, 011503 (2008).

\bibitem{hbchpt}
  E.~E.~Jenkins, A.~V.~Manohar,
  Phys.\ Lett.\  {\bf B255}, 558-562 (1991).

\bibitem{LSMsym1}
  T.~Hatsuda and M.~Prakash,
  Phys.\ Lett.\  B {\bf 224}, 11 (1989).

\bibitem{pisarski}
  D.~Zschiesche, L.~Tolos, J.~Schaffner-Bielich and R.~D.~Pisarski,
  Phys.\ Rev.\  C {\bf 75}, 055202 (2007).

\bibitem{LSMsym2}
  C.~Sasaki and I.~Mishustin,
  Phys.\ Rev.\  C {\bf 82}, 035204 (2010).

\bibitem{LSMasym}
  V.~Dexheimer, S.~Schramm and D.~Zschiesche,
  Phys.\ Rev.\  C {\bf 77}, 025803 (2008),
  V.~Dexheimer, G.~Pagliara, L.~Tolos, J.~Schaffner-Bielich and S.~Schramm,
  Eur.\ Phys.\ J.\  A {\bf 38}, 105 (2008).

\bibitem{gallas2011}
  S.~Gallas, F.~Giacosa and G.~Pagliara, ``Nuclear matter within  a dilatation-invariant parity doublet model:
  the  role  of the tetraquark at nonzero density,'' arXiv:1105.5003 [hep-ph].



\bibitem{weinberg}  S.~Weinberg,
  Phys.\ Rev.\ Lett.\  {\bf 65}, 1177 (1990).

\bibitem{georgi}  H.~Georgi,
  Nucl.\ Phys.\  B {\bf 331}, 311 (1990).

\bibitem{kuoetal} L.~W.~Siu, J.~W.~Holt, T.~T.~S.~Kuo and G.~E.~Brown,
  Phys.\ Rev.\  C {\bf 79}, 054004 (2009).

\bibitem{dklr} D. Dong, T.T.S. Kuo, H.K. Lee and M. Rho, to appear.

\bibitem{HBChPT}
  E.~E.~Jenkins, A.~V.~Manohar,
  Phys.\ Lett.\  {\bf B255}, 558-562 (1991).

\bibitem{PMR}
  T.~-S.~Park, D.~-P.~Min, M.~Rho,
  Phys.\ Rept.\  {\bf 233}, 341-395 (1993).

\end{thebibliography}
\end{document}